# Thermal properties of nanocrystalline silicon nanobeams

Jeremie Maire[1,2]*, Emigdio Chávez-Ángel[1], Guillermo Arregui[1,3], Martin F. Colombano[1,3], Nestor E. Capuj[4,5], Amadeu Griol[6], Alejandro Martínez[6], Daniel Navarro-Urrios[1,7], Jouni Ahopelto[8] and Clivia M. Sotomayor-Torres[1,9]

[1] Catalan Institute of Nanoscience and Nanotechnology (ICN2), CSIC and BIST, Campus UAB, Bellaterra, 08193 Barcelona, Spain.

[2] Institut de Mécanique et d'ingénierie (I2M), CNRS UMR5295, 33405 Talence, France

[3] Depto. Física, Universidad Autónoma de Barcelona, Bellaterra, 08193 Barcelona, Spain.

[4] Depto. Física, Universidad de La Laguna, 38200 San Cristóbal de La Laguna, Spain.

[5] Instituto Universitario de Materiales y Nanotecnología, Universidad de La Laguna, 38071 Santa Cruz de Tenerife, Spain.

[6] Nanophotonics Technology Center, Universitat Politècnica de València, 46022 València, Spain.

[7] MIND-IN2UB, Departament d'Enginyeria Electrònica i Biomèdica, Facultat de Física, Universitat de Barcelona, Martí i Franquès 1, 08028 Barcelona, Spain.

[8] VTT Technical Research Centre of Finland Ltd, P.O. Box 1000, FI-02044 VTT, Espoo, Finland

[9] Catalan Institute for Research and Advances Studies ICREA, 08010 Barcelona, Spain.

* Corresponding author. Email: jeremie.maire@u-bordeaux.fr
    Current address: I2M, CNRS UMR5295, 33405 Talence, France

**Abstract**

Controlling thermal energy transfer at the nanoscale has become critically important in many applications and thermal properties since it often limits device performance. In this work, we study the effects on thermal conductivity arising from the nanoscale structure of free-standing nanocrystalline silicon films and the increasing surface-to-volume ratio when fabricated into suspended optomechanical nanobeams. We characterize thermal transport in structures with different grain sizes and elucidate the relative impact of grain size and geometrical dimensions on thermal conductivity. We use a micro-time-domain thermoreflectance method to study the impact of the grain size distribution, from 10 to 400 nm, on the thermal conductivity in free-standing nanocrystalline silicon films considering surface phonon and grain boundary scattering. We find a drastic reduction in the thermal conductivity, down to values of 10 $W.m^{-1}.K^{-1}$ and below, which is just a fraction of the conductivity of single crystalline silicon. Decreasing the grain size further decreases the thermal conductivity. We also observe that this effect is smaller in OM nanostructures than in membranes due to the competition of surface scattering in decreasing thermal conductivity. Finally, we introduce a novel versatile contactless characterization technique that can be adapted to any structure supporting a thermally shifted optical resonance and use it to evaluate the thermal conductivity. This method can be used with optical resonances exhibiting different mode profiles and the data is shown to agrees quantitatively with the thermoreflectance measurements. This work opens the way to a more generalized thermal characterization of optomechanical cavities and to create hot-spots with engineered shapes at desired position in the structures as a means to study thermal transport in coupled photon-phonon structures.

**Keywords**

Thermal conduction; Optomechanics; phonons; silicon; nanostructure; polycrystalline; thermal characterisation methods

**Introduction**

Depending on the application, materials with widely varying thermal properties may be required. While for thermoelectric applications ultra-low thermal conductivity is essential[1,2], high thermal dissipation rates are mandatory for thermal management in microelectronics[3]. Therefore, it is crucial to understand thermal transport at the nanoscale to design structures with optimized thermal properties for a given application. Optomechanics (OM)[4] is one such application for which one typically wants to avoid absorption and the resulting heating, either to operate the resonator in the quantum regime, which is why two-dimensional structures are usually preferred over nanobeams, or even to induce amplification by dynamical back-action[5]. However, other approaches use thermal properties to control the mechanical resonator, e.g., self-pulsing-induced lasing [6–8] or bolometric back-action. Therefore, the direction towards thermal management in OM depends on the targeted application, making it essential to acquire direct information on thermal properties instead of having them as free parameters in complex models.

In the context of micro- and nanoscale thermal transport, single crystal silicon (c-Si) has been widely used as a platform to study thermal engineering with structures like phononic crystals. Si nanostructures exhibit strongly reduced thermal conductivity[9–14], non-diffusive thermal transport[15] and tuning of the thermal conductivity through the modified dispersion relation in phononic crystals[16,17]. In these structures, surface phonon scattering is the main mechanism impacting thermal properties when the dimensions are smaller than the mean free path of phonons in the bulk material, i.e., in the range of 100 nm to a few micrometers[18–22]. Nanocrystalline Si (nc-Si) is a specific type of polycrystalline Si in which the grain size is well below 1 μm. Due to the relatively easy tuning of the mechanical, optical, electrical and thermal properties by tailoring the stress and grain-size[23], as well as controlled material fabrication with conventional low temperature amorphous Si deposition techniques, nc-Si is widely used in MEMS offering a cost-competitive alternative to crystalline silicon in many practical scenarios such as 3D integration[24]. In nc-Si, scattering of phonons at grain boundaries adds to other phonon scattering mechanisms. This kind of scattering further reduces the thermal conductivity in nc-Si, cf. single crystal silicon, in nanostructures such as phononic crystals[25,26]. In this work, we quantify the relative effect of grain boundary scattering on thermal conductivity in corrugated OM nanobeams. Recently, nc-Si was shown to be an excellent, versatile and cost-competitive alternative to its crystalline counterpart to exploit OM non-linear dynamics under ambient conditions, with non-linear dynamic functions such as mechanical lasing and chaos, over a much larger frequency bandwidth compared to Silicon-on-Insulator (SOI) devices[27]. A complementary study of various properties of nc-Si thin films realised at different annealing temperatures focused on different grain sizes and tensile stress[23].

Here, we investigate the thermal properties of nc-Si membranes and optomechanical nanobeam cavities with different grain sizes, which are structurally characterized by image processing of dark-field transmission electron microscope (TEM) images. To access the thermal decay rates and conductivity, we use two pump-probe techniques. For a direct assessment we use micro time-domain thermoreflectance (μ-TDTR) as a proven technique and extract the thermal conductivity by fitting the experimental temperature rise with finite element modelling (FEM). We demonstrate that the thermal conductivity of the nanocrystalline membrane is at least 4 times lower than its crystalline counterpart, with a further reduction by a factor 2 when grain size decreases to an average of 163 nm. A similar phenomenon occurs in in the nanobeams studied here but its relative impact is smaller than in membranes as it competes with enhanced surface scattering stemming from the lower dimensions of these nanobeams. We further introduce a new two-laser measurement technique for direct measurements of thermal properties of in operandi OM nanobeams, similar to time-stretch spectroscopy used for microspheres[28]. This technique is based on the cooling rate of an optical resonance to measure the thermal decay rate of the cavity. We unveil the potential of this technique, which can be readily applied in current OM devices with an optical cavity, and compare the results to those obtained by μ-TDTR. We show how optical resonances with different profiles can induce a thermal dissipation rate variation of up to 20% in a single optical cavity. Our results provide new insights on the thermal characteristics of nc-Si devices, which may find application in optics, optomechanics and energy-harvesting devices.

**Nanocrystalline silicon films and structural analysis**

The membranes and nanobeams were fabricated on wafers with a thick silicon dioxide film and a 220 nm thick nc-Si layer on top of the oxide, resembling the SOI wafers typically used in fabrication of optomechanical devices. The wafers were produced by the following process. A thick $SiO_2$ layer is first grown by wet oxidation at 1050 °C on a Si wafer, followed by a 220 nm thick layer of amorphous Si (a-Si) deposited at 574 °C by low pressure chemical vapour deposition (LPCVD). Four wafers were annealed at different temperatures: 650°C (OMS1), 750°C (OMS2), 850°C (OMS3) and 950°C (OMS4) for 60 min. The annealing converts the amorphous Si to nc-Si, with the grain size distribution ranging from a few nm to a few hundreds of nm. The thickness of the nc-Si films after annealing was 211 nm measured by spectroscopic reflectometry.

Bright field planar TEM images of the nanocrystalline film annealed at 950°C (OMS4) and the corresponding selective area diffraction pattern are shown in Figure 1. The crystallites do not exhibit a preferential orientation. The image is representative of all the samples used in this work with only the grain size distribution obtained by analysing dark field TEM images varying with annealing temperature. In all samples the crystallite size has a log-normal distribution with the average crystallite size of 163 nm (OMS1), 171 nm (OMS2), 187 nm (OMS3), and 215 (OMS4). The a-Si layer is under compressive stress after deposition and annealing converts it to tensile, measured to be 290 MPa (OMS1), 250 MPa (OMS2), 170 MPa (OMS3) and 90 MPa (OMS4). Further details of the nc-Si wafer fabrication, grain size and stress analysis can be found in Ref. [23].

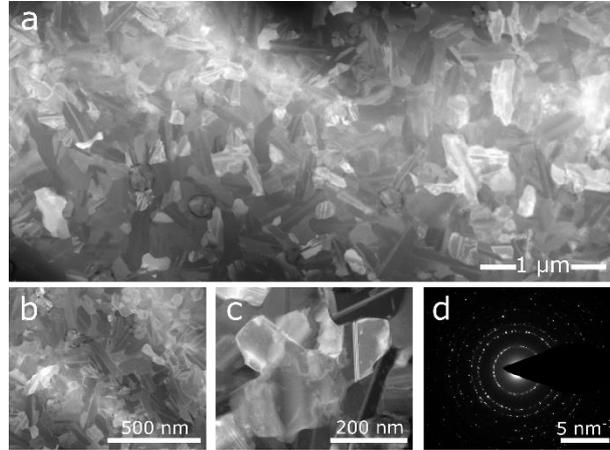

**Figure 1. Structure of nanocrystalline silicon. a-c.** Bright field planar TEM images of the nc-Si film with an average grain size of 215 nm and $T_a$ = 950°C with increasing zoom, showing the randomly oriented grains and their relatively large size distribution. **d.** Selective area electron diffraction image.

To identify the effects arising from the nanocrystallinity of the material, identical reference samples were fabricated on commercial SOI wafers with a 220 nm thick single crystalline Si layer. A summary of the structural characteristics of all samples is given in Table 1 and samples will be designated by their average grain size from now on.

**Table 1.** Summary of structural characteristics of fabricated samples

|      | Thickness (nm) (monitor wafers) | $T_a$ (°C) | Average grain size (nm) | Tensile stress (MPa) |
|------|---------------------------------|------------|-------------------------|----------------------|
| **c-Si**  | 220 |     |         | - 39 |
| **OMS4**  | 211 | 950 | **215** | 90   |
| **OMS3**  |     | 850 | **187** | 170  |
| **OMS2**  | 211 | 750 | **171** | 250  |
| **OMS1**  |     | 650 | **163** | 290  |

**Thermal conductivity measurements by µ-TDTR**

The thermal properties of a device are affected by the structure of the material itself, the processing steps, the operating conditions and the geometry of the device. It has been shown that the surface can play a major role in thermal conductivity when the surface-to-volume ratio (S/V) is large[9]. We use micro time-domain thermoreflectance technique (µ-TDTR) to measure the thermal conductivity of the nanocrystalline films and the nanobeams. The thermal conductivity extracted with this technique corresponds to the ability of the material to conduct heat by accounting for the effects of nanopatterning, such as phonon boundary scattering and the reduction in the volume of material. µ-TDTR has been previously used to characterize the thermal properties of numerous nanostructures, including membranes, nanobeams[29,30], phononic crystals[31–35] or phonon lenses[36], and to demonstrate effects such as heat focusing[36], ballistic thermal transport[30,37,38] and to highlight the contribution of the wave nature of phonons to thermal transport at cryogenic temperatures[17]. We use two different sample designs shown in Figure 2. One for investigation of the role of the nanocrystalline material and the other to investigate the role of surfaces. The gold pad in the middle of the samples acts as a transducer and as the detector. The temperature of this pad is directly related to its reflectivity, which is probed by a continuous-wave laser (532 nm). A 405 nm pulsed laser periodically heats the metal and lets it cool down between pulses. The heating time is chosen to allow the system to reach the steady state. The temperature gradient across the structures then progressively disappears as the heat flows from the central pad to the heat bath. The characteristic time for the temperature gradient to vanish is measured and its inverse gives the heat dissipation rate. The cooling curve can be described with a single-parameter exponential decay in time, $\exp(-\gamma t)$, where $\gamma$ is the heat dissipation rate of the system. Further details about the technique and a schematic of the setup are given elsewhere[17,36].

The nanobeams consist of an optomechanical cavity flanked on both sides by an OM Bragg mirror to prevent the leakage of the colocalized optical and mechanical modes. Each part of the OM nanobeam consists of a repetition of unit cells comprising a central beam with stubs on both sides and a cylindrical hole in the center. The OM nanobeam is shown in Figure 2. The OM cavity used to confine optical and GHz mechanical modes corresponds to the central region of 12 unit cells in which the pitch, hole diameter and stub width along the beam axis progressively decrease towards the center by a factor Γ, which is typically around 0.8. On both sides of this region, the Bragg mirrors consist of 10 unit cells. The nominal values of the pitch, hole diameter and stub width are 500, 300 and 250 nm, respectively. The fabrication process of these OM nanobeams is detailed elsewhere [8]. The fabrication process is explained in the Supporting Information.

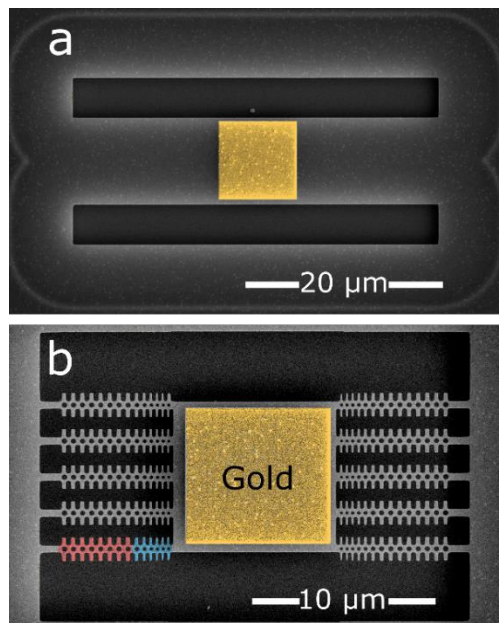

**Figure 2. Structures for μ-TDTR measurements. a.** SEM image of a suspended nc-Si membrane. **b.** SEM image of the nanobeams. The Bragg mirror of the bottom left beam is shown in red and half of the optomechanical cavity in blue. The central gold pad is the transducer for μ-TDTR measurements.

We first investigate the effect of nanocrystals on thermal transport by measuring nominally identical suspended membranes fabricated on the wafers with different annealing temperatures $T_a$. In these structures, the only phonon scattering mechanism that is not intrinsic to the material is surface scattering at the top and bottom surfaces of the membranes. As this scattering only depends on the spacing between the surfaces, which is identical for all samples, and the surface quality, the differences in heat dissipation rates solely stem from the crystallinity. For each of the nanocrystalline samples, we measure three nominally identical structures. We then calculate the average value of the heat dissipation rate $\gamma$ and the standard deviation of the measurements gives the error bars. The results are shown in Figure 3a. It is clear in the figure that $\gamma$ increases with annealing temperature and it is not an intrinsic feature instead directly depending on the geometry, as can be seen that the shorter membranes dissipate heat faster. The intrinsic thermal property is the thermal conductivity. Due to the geometry of the structure, neither the 1D nor the 2D heat equation can be used to analytically deduce the thermal conductivity. Hence, we use FEM simulations to virtually reproduce our µ-TDTR experiments, with a heating phase modelled by an inward heat flux in the metal pad, of same duration as the experimental heating, and a cooling phase. The temperature is "probed" at the center of the metal pad. In this 3D FEM model, the thermal conductivity is the only free parameter. The heat dissipation rates obtained for different values of the thermal conductivity are then compared with the experimental dissipation rate to extract the experimental thermal conductivity. The uncertainty on the measurement of the structure dimensions (±3 nm) results in an error in thermal conductivity of less than 5%. Further details of the measurement system and FEM simulations are provided in [17,36,39].

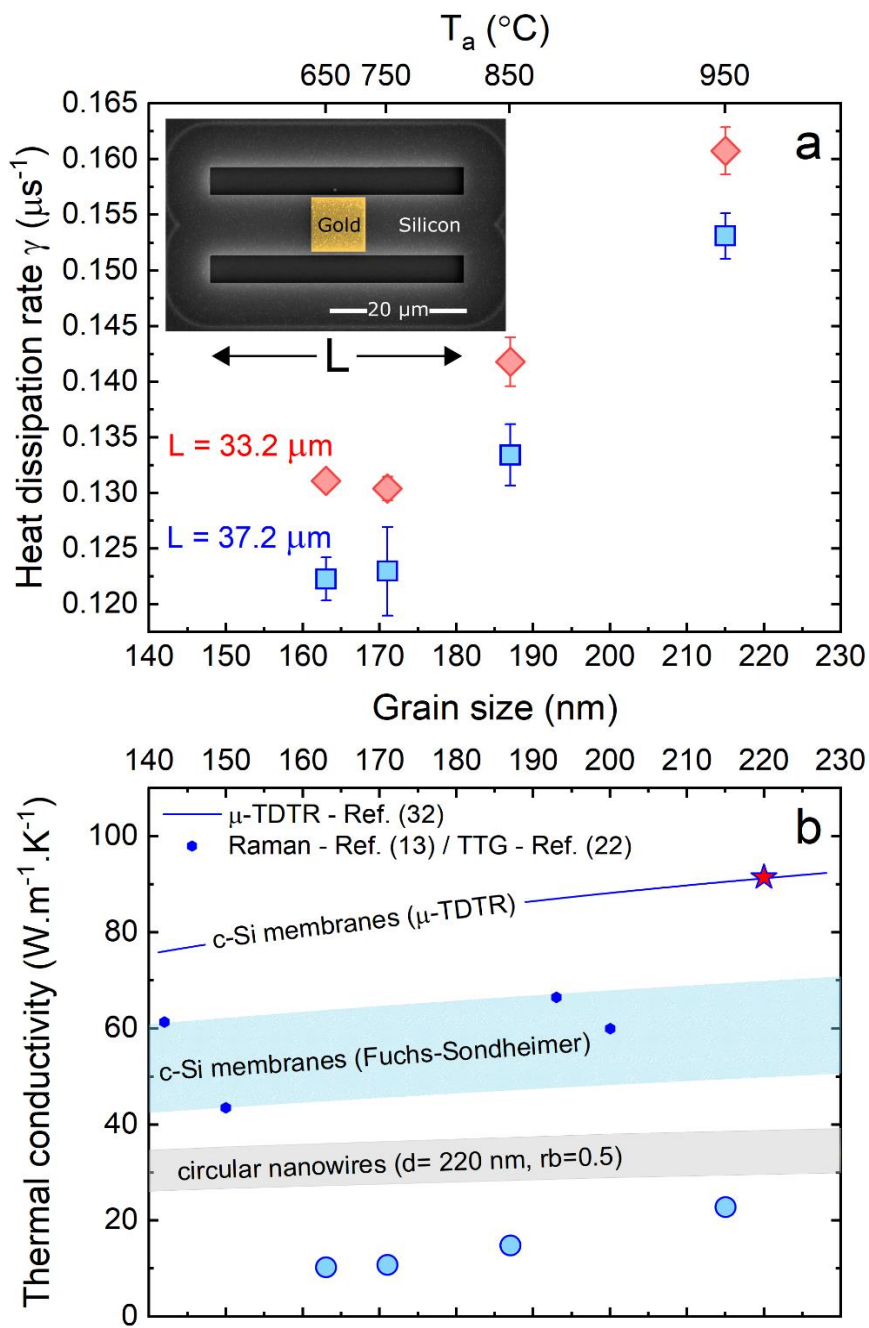

**Figure 3. Thermal properties of nanocrystalline Si membranes. a** Heat dissipation rate $\gamma$ measured as a function of the annealing temperature $T_a$ and the corresponding average grain size for two membrane lengths L. **b** Thermal conductivity of the same membranes as a function of $T_a$, shown as large light blue dots. The values range from 10 to 20 $W.m^{-1}.K^{-1}$ and correspond to the average for each grain size of the data from panel a. Error bars are contained within the circles. The blue "c-Si membranes" line corresponds to a fit of the µ-TDTR data for crystalline silicon, adapted from Ref. [33]. The red star indicates the value of thermal conductivity of a 220 nm thick c-Si membrane estimated from that data. The small blue dots and the blue stripe represent the thermal conductivity of single crystalline membranes measured by Raman thermometry [13] and the transient thermal grating (TTG) technique [22]. The "grain size" is taken to correspond to the thickness of the membranes, and calculations are based on the Fuchs-Sondheimer model and consider the different amount of impurities in the films. The grey stripe corresponds to similar calculations for circular nanowires of diameter 220 nm as a function of the grain size for a fixed carrier-reflection parameter Rb = 0.5 using a Mayadas model. A detailed explanation of the calculations is given in the Supporting Information.

The results of that analysis as a function of the grain size are shown in Figure 3b. For comparison, the values for c-Si are given, with the x-axis representing the membrane thickness, which is the limiting dimension in that particular case. For each average grain size, the thermal conductivity value is a weighted average of the data displayed in Figure 3a. Data from c-Si membranes measured by µ-TDTR[33], Raman thermometry[13] and transient thermal gratings (TTG)[22] yield thermal conductivity between 40 and 90 W.m$^{-1}$.K$^{-1}$ for the dimensions covered by this study. Due to fabrication limitations related to buckling of single crystal silicon membranes, we could not measure their thermal conductivity. Nonetheless, the Fuchs-Sondheimer model has been extensively used to describe that effect and it shows extremely good agreement with the reported values in the literature[37]. In such membranes, thermal transport stills occur in two directions. We then calculate the thermal conductivity in nanowires of diameter 220 nm using a Mayadas model, in which thermal transport is one-dimensional. We also introduce grains in the simulated nanowires, with a fixed carrier-reflection parameter Rb=0.5. In Fig. 3b, we see that the expected thermal conductivity is indeed lower in these nanowires than in the membranes. Details of these calculation are given in Supplementary Information. In the nc-Si membranes studied in this work, thermal transport is impeded in all three directions. We observe that even when the average grain size of 215 nm ($T_a$ = 950°C) of these membranes is similar to the thickness, the thermal conductivity is reduced by 54% compared to the single crystal case. This reduction increases to 77% in the sample with the smallest average grain size of 163 nm. To explain these results, it is important to note that although specular phonon scattering events can occur at atomically flat surfaces at room temperature[40], thermal transport is considered diffusive overall[22]. The difference in thermal conductivity between crystalline and nanocrystalline silicon membranes therefore mainly stems from the scattering events at grain boundaries, whose frequency of occurrence in nc-Si is directly linked to the grain size distribution. Furthermore, a recent experimental study on 145 nm-thick crystalline Si membranes[41] shows that phonons with mean free path above 215 nm contribute nearly 20% to thermal conductivity at room temperature. This proportion is expected to be relatively higher in our membranes due to the increased thickness and subsequent shift of the mean free path distribution towards higher values. Anufriev et al. [40] showed that phonon mean free path smaller than 400 nm contribute significantly to thermal conductivity in a 145 nm thick membrane. This suggests that the grain size distribution in our samples cover the range of mean free paths with the strongest contribution to thermal conductivity. It explains the strong suppression of the thermal conductivity measured here. Furthermore, we see that our experimental values of thermal conductivity lie below the expected values for nanocrystalline nanowires from the Mayadas model with a carrier reflection parameter of 0.5. Although no direct quantitative comparison is possible between these two sets of values, they suggest that the transmission at grain boundaries in our samples might be lower than the value input in the model. High spatial resolution measurements on single grain boundaries might shed light on the transmission at single grain interfaces.

The role of the S/V ratio using nc-Si optomechanical nanobeams is shown in Figure 4 where the same trend observed in nc-Si membranes is qualitatively observed also for OM nanobeams, that is, the heat dissipation rate decreases with increasing length and with decreasing grain size. In the case of periodic nanobeams, Γ corresponds to the decrease of the geometrical parameters towards the central island of the OM structure (see Figure 8 in the Supporting Information). In our experiment, the measured values of heat dissipation rate γ are insufficiently sensitive to Γ. Therefore, no distinction is made in the rest of this work between structures with differing Γ values. The inset to Figure 4 shows that the grain size impacts heat dissipation in a similar way as in membranes, with an increase of the heat dissipation rate by more than 30% comparing samples made out of nc-Si annealed at $T_a$ = 650°C and $T_a$ = 950°C. However, the difference with c-Si is much smaller than in the suspended membranes since the heat dissipation is ~19% slower in nc-Si nanobeams with $T_a$ = 950°C than in identical nanobeams made of c-Si. This phenomenon is attributed to the fact that the heat dissipation is already strongly suppressed in c-Si nanobeams due to surface phonon scattering.

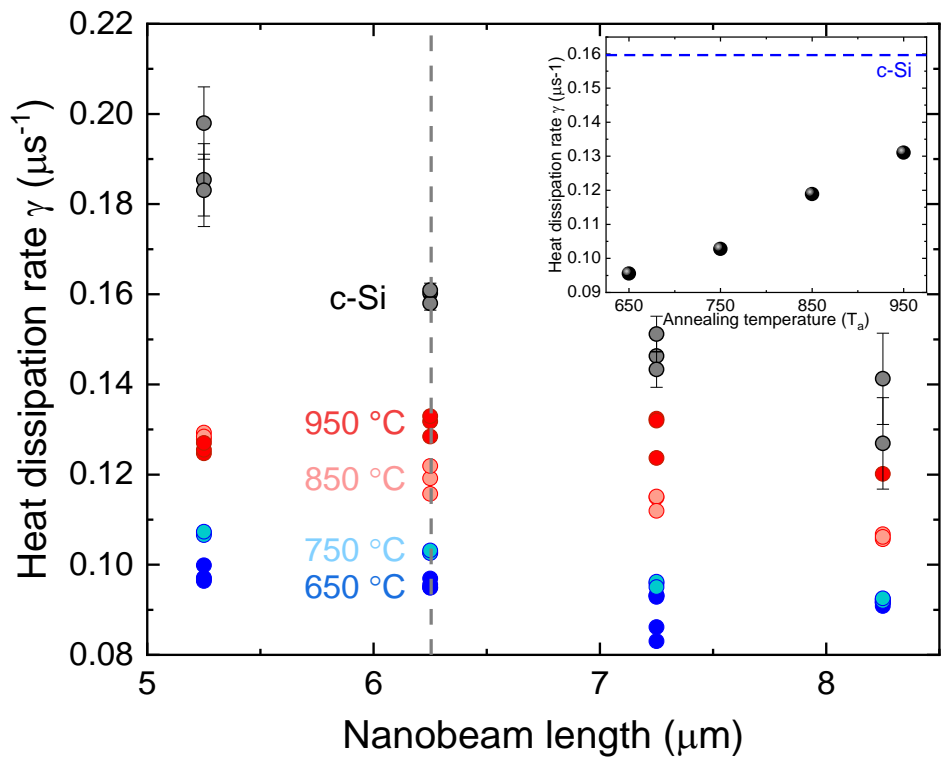

**Figure 4. Impact of nanostructuring and grain size on the heat dissipation rate.** Heat dissipation rate measured with the μ-TDTR technique as a function of nanobeam length and annealing temperature $T_a$. The grey dots show the rate measured from geometrically identical single crystalline nanobeams. **Inset.** Average heat dissipation rate of 6.25 μm long nanobeams as a function of the annealing temperature. Errors bars are shown with the data from SOI samples only for clarity and correspond to the standard deviation of the measurement in structures of identical length.

Due to the multi-parameters involved, evaluating the impact of geometry on the thermal properties is not straightforward. Two main parameters have been extensively used in the literature as a way to encompass the effect of geometry in nanostructures and phononic crystals, namely, the neck size – the smallest width available for phonons to travel through – and the S/V ratio to highlight the importance of surface scattering. We varied the geometry for each sample by increasing the hole diameter and reducing the width of the central beam as well as the width and length of the stubs. These changes effectively increase the S/V ratio and are schematically shown in the upper part of Figure 5. The data confirms that for a given S/V ratio, that is, for a similar geometry, the heat dissipation rate increases with increasing grain size but remains lower than that of c-Si. It is interesting to note that for the highest S/V ratio, which corresponds to the smallest neck, heat is dissipated nearly as fast in c-Si as it is in the sample with $T_a$ = 950°C. This observation suggests that as surface scattering increases and the neck becomes small enough – below 75 nm in our samples – the impact of crystallinity becomes negligible for that sample. The second observation is highlighted by the coloured areas in Figure 5. Note that these coloured areas are visual guides and not fits to the data. Given that heat is dissipated through the nanobeams but also through air, the slope of each area qualitatively indicates the relative impact of geometry on heat dissipation. We see in Figure 5 that nanocrystalline material has less dependence on the S/V ratio compared to single crystal silicon and the geometry seems to have a lower impact as the grains become smaller.

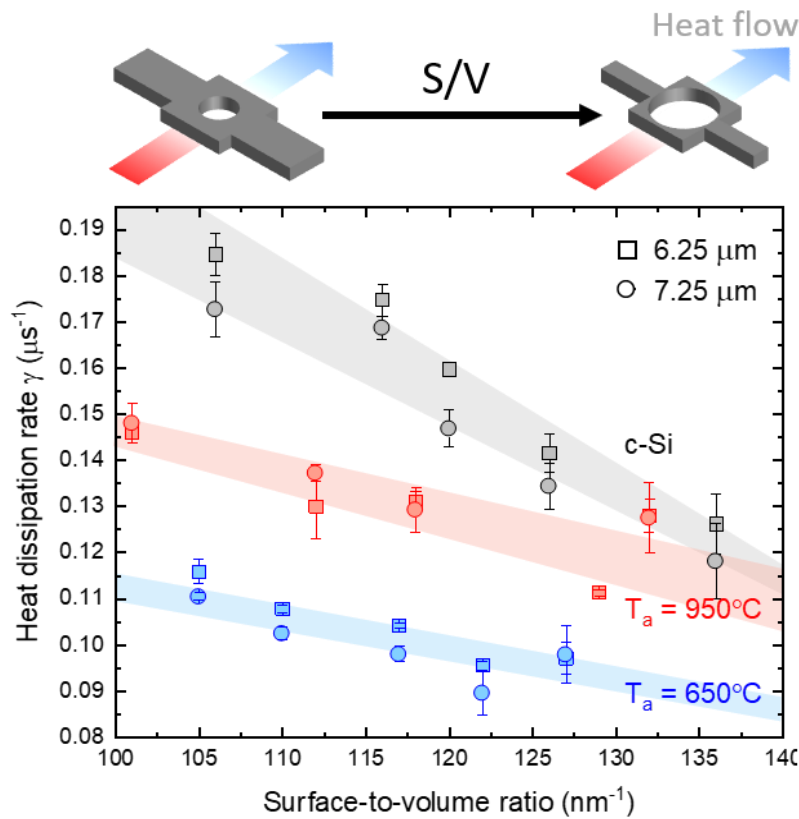

**Figure 5. Impact of S/V ratio on heat dissipation rate.** Heat dissipation rate as a function of the S/V ratio in nanobeams made of nc-Si with average grain size of 163 nm ($T_a$ = 650°C, blue), 215 nm ($T_a$ = 950°C, red) and of single crystalline silicon c-Si (grey). Data is shown for two nanobeam lengths: 6.25 µm (squares) and 7.25 µm (circles). The coloured stripes are guide to the eye.

We summarize the µ-TDTR data in Table 2. The data obtained from membranes is compared to its crystalline counterpart estimated from µ-TDTR in the literature, whereas the data from nanobeams is normalized to that of the corresponding membrane and to that of the c-Si nanobeam of similar S/V ratio. An increase of the S/V ratio corresponds to an overall decrease in the nanobeam neck, stub width and depth, and an increase in hole diameter. The data for nanobeams is obtained for a length of 6.25 µm as shown in the inset to Figure 4. The thermal conductivity data for nanobeams with average grain size of 187 nm is highlighted since it will be compared to values obtained by the optical resonance cooling technique.

**Table 2.** Summary of thermal characteristics of membranes and nanobeams.

| | Membranes | | Nanobeams | | | | | |
| | | | S/V= 116.4 ± 0.8 nm$^{-1}$ | | | S/V= 127.9 ± 1.2 nm$^{-1}$ | | |
| | K (W.m$^{-1}$.K$^{-1}$) | K/K$_{c-Si}$ | K (W.m$^{-1}$.K$^{-1}$) | K/K$_{c-Si}$ | K/K$_{membrane}$ | K (W.m$^{-1}$.K$^{-1}$) | K/K$_{c-Si}$ | K/K$_{membrane}$ |
|---|---|---|---|---|---|---|---|---|
| **c-Si** | 91.5 | 1 | 42.73 | 1 | 0.47 | 25.33 | 1 | 0.28 |
| **215 nm** | 22.8 | 0.249 | 16.06 | 0.38 | 0.70 | 8.71 | 0.34 | 0.38 |
| **187 nm** | 14.9 | 0.163 | **9.63** | 0.23 | 0.65 | **6.43** | 0.25 | 0.43 |
| **171 nm** | 10.9 | 0.119 | 5.95 | 0.14 | 0.55 | 2.09 | 0.08 | 0.19 |
| **163 nm** | 10.3 | 0.113 | 5.27 | 0.12 | 0.51 | 3.22 | 0.13 | 0.31 |

**Thermal dynamics by optical resonance cooling**

Since the heat dissipation rate $\gamma$ obtained by µ-TDTR is not strictly equivalent to the thermal decay rate $\Gamma_{th}$ in optomechanical devices due to the added contribution of the suspended central section and gold pad and their heat capacity. The nanobeams studied here have been extensively used as a platform for OM experiments, including self-pulsing[6,8], chaotic behaviour[7], injection locking[42] and synchronisation[43], among others[23,27]. In this context, non-linear dynamics involving self-pulsing limit cycle has played a key role. Such non-linear dynamics involve not only free carrier dispersion but also the temperature increase of the cavity, which directly depends on the thermal properties of the nanostructure. Of particular interest is the thermal decay rate $\Gamma_{th}$. This thermal decay rate is used in modelling to explain the dynamics of the system[6], but so far has only been estimated. Here, we demonstrate a way to directly measure $\Gamma_{th}$ using the cooling rate of the optical resonance of an OM cavity. Details of the method are given in the Supporting Information. Briefly, two tuneable laser beams, with polarization states independently controlled, are multiplexed into a tapered fiber. The thinnest part of the taper is twisted into a loop, which is then positioned parallel to the nanobeam as shown schematically in Figure 6a. The distance between the fiber and the nanobeam is approximately 0.2 $\mu m$ so that the long tail of the evanescent field of the fiber mode locally excites the resonant optical modes of the cavity. The light travelling through the fiber from either laser can couple to the optical cavity modes if its wavelength corresponds to one such resonance. When measurements are performed in transmission, the light that was not coupled to the resonance continues its path through the optical fiber. The signal is then separated so that each branch is dedicated to one laser. The signal in each branch goes through an in-line bandpass filter adjusted to one of the two lasers and is then detected by a fast InGaAs photodetectors (12 GHz bandwidth). The signals from both detectors are then transmitted to and recorded by an oscilloscope. All measurements are performed in an anti-vibration cage at ambient temperature and atmospheric air pressure. More details of the measurement system to characterise optomechanical cavities are given in a previous work[43] and a schematic of the setup is shown in Supplementary Information.

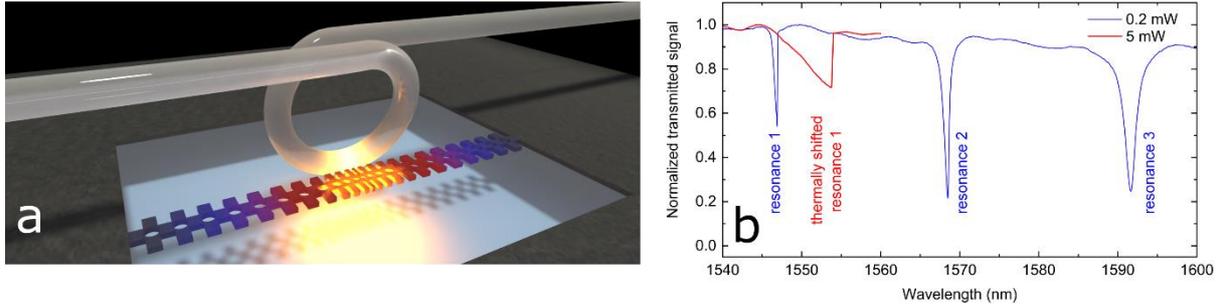

**Figure 6. Principle of optical resonance cooling. a** Schematic of the experimental configuration involving the fibre and the nanobeam. Two near-infrared lasers with wavelengths around 1.5 µm are evanescently coupled to the optomechanical cavity of the nanobeam via a tapered optics fiber loop. One laser induces a temperature distribution in the nanobeam, represented by the red to blue gradient along the beam, whereas the second laser, represented in yellow, probes the delay of passage of the optical resonance at given wavelength. **b** Optical transmission spectra through the optical fiber in the sample with average grain size 187 nm. The transmission measured at low power is shown in the blue trace. The power is selected to avoid heating the cavity modes. The three optical resonances studied in this work are detected. Heating of the cavity at higher laser power (5 mW) is shown in the red trace, with the first optical resonance thermally shifted to longer wavelength. The spectra are obtained by sweeping the pump laser from short to long wavelengths. The resonance cooling data shown in Figure 7 is measured by tuning the probe laser between the non-thermally shifted position of the resonance and its position thermally shifted by 4.5 nm by the 5.5 mW pump laser.

To obtain the thermal decay rate $\Gamma_{th}$ we rely on a pump-probe technique. The first three cold resonances of the measured structure are shown in Figure 6b with the blue line. First, we turn on the pump laser in a blue-detuned position, at shorter wavelength compared to the empty "cold" cavity resonance, i.e., the position of the resonance without photons in the cavity, which corresponds to the 1st resonance at 1545.5 nm in Fig. 7. The wavelength of the pump laser is then continuously increased, and as this pump laser couples to the resonance in the cavity, the transmission through the fiber exhibits a dip. At fixed laser power, we continue increasing the wavelength of the pump laser, thus continuously shifting the resonance until it is approximately 4.5 nm away from its "cold" position while verifying that the cavity remains below the self-pulsing limit. The spectrum of a thermally shifted resonance is shown in red in Figure 6b. The probe laser is then turned on at a wavelength between the initial position of the resonance measured at low laser power and the position thermally shifted by 4.5 nm by the high-power pump laser, and the wavelength-filter is adjusted to the probe laser. A waveform generator then gives an impulse to the pump laser, switching it off, which triggers the acquisition of the waveform of the probe laser on the oscilloscope. This cycle is repeated for each new wavelength of the probe laser, effectively mapping the spectrum between the cold and heated wavelengths of the optical resonance. The recorded curves, examples of which are displayed in Figure 7b, are then fitted with a Lorentzian function to identify the transmission minimum. The wavelength of the probe laser is converted to a temperature rise using the thermo-optic coefficient. This coefficient gives the shift of the optical resonance with temperature increase and was measured to be 0.09 nm.K$^{-1}$ in our previous work[27]. The temperature rise is then transformed into the time elapsed after switching off the pump laser, as shown in Figure 7, and fitted with an exponential decay function. The extracted decay rate then corresponds to the thermal decay rate $\Gamma_{th}$. To extract the thermal conductivity, we use finite element method (FEM) simulations by importing the nanobeam geometry from an SEM image and reproducing the experimental situation in a 3D FEM model. We first calculate the optical resonances of the optical cavity and use it as a heater. Once the temperature in the cavity has reached a steady-state, we turn off the heater and monitor the cooling rate of the nanobeam. The simulation is repeated for different values of the thermal conductivity of the nanobeam and the extracted decay rates are matched to the experimental value to identify the real thermal conductivity. Additional details about the setup and the thermal conductivity extraction are given in the Supplementary Material.

Figure 7 shows the results of the measurements of the first optical resonance of an OM nanobeam made out of the nc-Si with average grain size 187 nm ($T_a$ = 850 °C). The cavity and Bragg mirrors comprise of 6 and 10 cells on each side, respectively, resulting in a nanobeam length of 16.5 µm. This measurement was repeated for the 2$^{nd}$ and 3$^{rd}$ optical resonance of the same cavity. The three resonances display thermal decay times of 1.40, 1.67 and 1.61 µs, respectively. The different decay times are attributed to differences in the spatial mode profile of each of the optical resonances, as shown in the Supporting Information. The extracted thermal conductivity for each of the optical resonance is then 7.71, 7.04 and 7.56 W.m$^{-1}$.K$^{-1}$, respectively, giving an average thermal conductivity of 7.44 ±0.29 W.m$^{-1}$.K$^{-1}$. Although no direct quantitative comparison with µ-TDTR measurements is possible due to differences in geometry, namely, the absence of the central island and transducer, the thermal conductivity falls within the range obtained by the µ-TDTR method for similar geometries (see Table 2). Note that the extracted thermal conductivity considers the exact geometry, including the presence of the optical fiber next to the nanobeam. In optomechanics experiments, the optical fiber plays an important role in dissipating heat, which can be observed, e.g., in the change of the self-pulsing frequency mentioned above. Indeed, the volume of the fiber being much larger than that of the nanobeam compensates for the low thermal conductivity of glass, at 1.4 W.m$^{-1}$.K$^{-1}$, especially when considering nc-Si nanobeams with low thermal conductivity. The optical fiber therefore constitutes one of the main factors limiting the sensitivity of the technique in extracting the thermal conductivity of the nanobeams. The measurements presented here nonetheless show the validity of this method to determine thermal conductivity but more importantly to obtain the thermal decay rate associated to OM cavities in real characterisation configuration.

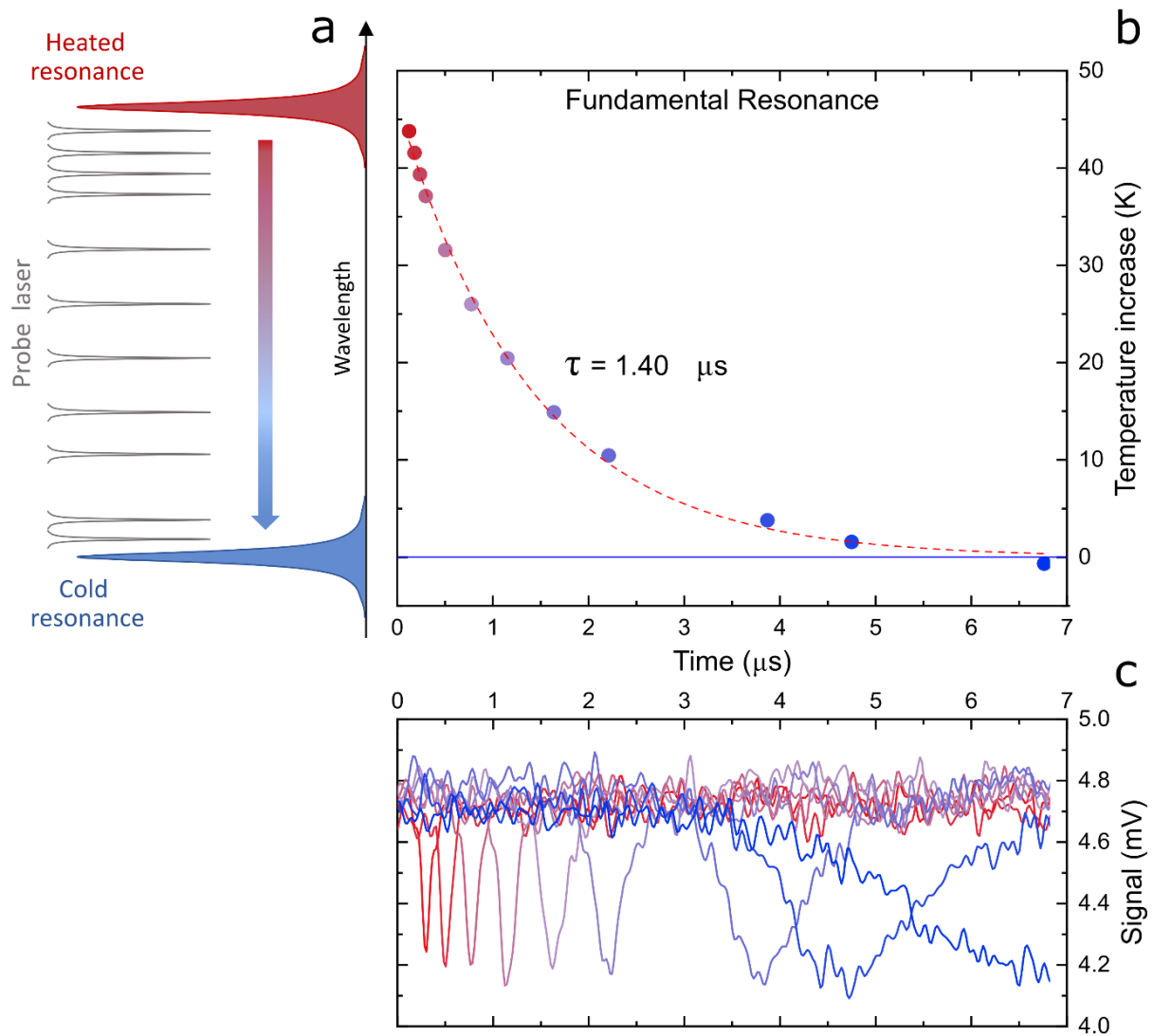

**Figure 7. Thermal measurements using optical resonance cooling. a** Schematic of the different wavelengths of the probe laser in-between the cold and hot resonance. **b** The optical resonance is heated by a pump laser, increasing the resonance wavelength. The pump is then switched off and the cavity cools down. A probe laser is set at different wavelengths but between resonance wavelengths of the heated and cold cavity. We record the time between the turning off of the pump laser and the dip corresponding to the passing of the resonance through the probe laser wavelength. The curve representing the probe wavelength as a function of time directly gives the cooling speed of the optical resonance and prior calibration of the resonance wavelength as a function of temperature gives the decay curve on the right. **c** Oscilloscope-recorded optical signal for different probe laser wavelengths. As the wavelength gets further away from the heated resonance wavelength, i.e., as the resonance cools down, the dip in the signal gets broader and occurs at a later time, denoting a slower and delayed passage of the probe signal through the optical resonance in the cavity.

**Discussion and conclusions**

In summary, we have used several different experimental techniques to investigate the structural and thermal properties of OM crystal cavities made of nanocrystalline silicon. Dark-field TEM analysis shows that different annealing temperatures result in different grain size distributions, which affect the thermal properties of the material. We used micro-time-domain thermoreflectance to measure heat dissipation rates in the nanocrystalline thin films and observe a strong reduction compared to the single crystalline films. This is attributed to phonon scattering at the grain boundaries. A recent study on c-Si membranes[41] suggest that the grain size distribution in our samples covers the range of mean free paths that contribute most to thermal conductivity, thus potentially explaining the low measured values of thermal conductivity. Measurements of the mean free path distribution in our nc-Si membranes should be the focus of a future study to quantify this aspect. We then investigated the thermal properties of released OM nanobeams. In such structures, the geometrical dimensions are small with respect to phonon mean free path, which means that surface scattering has a strong effect on thermal conductivity. We showed that the thermal conductivity is very low in nanocrystalline nanobeams, well below 10 $W.m^{-1}.K^{-1}$, and depends on the grain size. However, the impact of the grain boundaries in nanobeams is lower than in membranes as the grain boundary scattering competes with the surface scattering stemming from nanostructuring. Finally, we have developed a novel technique to directly measure thermal decay in nanostructures with an optical cavity. The method can be readily adapted to existing OM structures. By using a pump-probe technique to measure the cooling rate of a localized optical resonance, we were able to extract thermal decay rates in nanobeams without any modification of the structure. The values obtained are consistent with those from the thermoreflectance technique and show the strong potential of this novel contactless method for all nanostructures with optical cavities. The results presented in this work clarify the impact of grain boundaries on thermal transport in silicon micro- and nano-structures, in a context in which crystalline state is a key parameter in NEMS and/or NOEMS. Moreover, the versatility of the new characterization technique paves the way to more standardized measurements of thermal properties and to potentially study the impact of strain on thermal and optomechanical properties.

## Acknowledgements

This work was supported by the European Commission FET Open project PHENOMEN (G.A. Nr. 713450). ICN2 is supported by the S. Ochoa program from the Spanish Research Agency (AEI, grant no. SEV-2017-0706) and by the CERCA Programme / Generalitat de Catalunya. ICN2 authors acknowledge the support from the Spanish MICINN project SIP (PGC2018-101743-B-I00). DNU and MFC acknowledge the support of a Ramón y Cajal postdoctoral fellowship (RYC-2014-15392) and a Severo Ochoa studentship, respectively. ECA acknowledges financial support from the EU FET Open Project NANOPOLY (GA 289061). A. M. acknowledges support from Ministerio de Ciencia, Innovación y Universidades (grant PGC2018-094490-B, PRX18/00126) and Generalitat Valenciana (grants PROMETEO/2019/123, and IDIFEDER/2018/033).